\documentclass[aps, preprint, groupedaddress]{revtex4}

\usepackage{amsfonts, amsbsy, amssymb, amsmath, graphicx,  float} 

\newcommand{\pd}[2]{\frac{\partial {#1}}{\partial {#2}}}
\newcommand{\calK}{\mathcal{K}}
\newcommand{\calH}{\mathcal{H}}

\newcommand{\qb}{\bar{q}}
\newcommand{\pb}{\bar{p}}
\newcommand{\fb}{\mathfrak{B}}

\begin{document}


\title{Impenetrable Barriers in Phase Space for Deterministic Thermostats}



\author{Gregory S. Ezra}
\email[]{gse1@cornell.edu}
\affiliation{Department of Chemistry and Chemical Biology\\
Baker Laboratory\\
Cornell University\\
Ithaca, NY 14853\\USA}

\author{Stephen Wiggins}
\email[]{stephen.wiggins@mac.com}
\affiliation{School of Mathematics\\University of Bristol\\Bristol BS8 1TW\\United Kingdom}


\date{\today}

\begin{abstract}

We investigate the relation between the phase space structure of Hamiltonian and
non-Hamiltonian deterministic thermostats.
We show that phase space structures governing reaction dynamics in Hamiltonian systems map 
to the same type of phase space structures for the non-Hamiltonian isokinetic equations of 
motion for the thermostatted Hamiltonian. 
Our results establish a framework for analyzing thermostat dynamics using 
concepts and methods developed in reaction rate theory.

\end{abstract}


\maketitle


\section{Introduction}

\label{sec:introduction}

Deterministic thermostats are widely used to simulate equilibrium physical systems 
described by ensembles other than microcanonical (constant energy and volume, $(E, V)$),  such as constant 
temperature-volume $(T, V)$ or temperature-pressure $(T, p)$ 
\cite{Nose91,Morriss98,Leimkuhler04a,Hunenberger05,Bond07}.
Deterministic thermostats 
are typically obtained by augmenting the phase space variables of the physical system of interest 
with a set of additional variables
whose role is to alter the standard Hamiltonian system dynamics in such a way that
a suitable invariant measure in the system phase space is preserved.
In the familiar Nos\'{e}-Hoover (NH) thermostat \cite{Nose84,Hoover85}, for example, 
the exact dynamics preserves both
an extended energy $\calH$ and a suitable invariant measure, ensuring that, provided the extended system dynamics
is efectively ergodic on the timescale of the simulation, the physical system will sample its phase space
according to the canonical (constant $T$) measure.

Extended system thermostat dynamics can be either Hamiltonian \cite{Nose84,Dettmann96,Dettmann97,Dettmann99,Bond99} 
or non-Hamiltonian \cite{Hoover85,Evans90,Tuckerman99,Tuckerman01,Sergi01,Sergi03,Ezra04,Tarasov05,Ezra06,Sergi07}.  
An important motivation for the formulation of
Hamiltonian deterministic thermostats such as the Nos\'{e}-Poincar\'{e} system \cite{Bond99}
is the possibility of using
symplectic integration algorithms to compute trajectories \cite{SanzSerna94,Hairer02,Leimkuhler04a}.  

In this  approach, 
an extended Hamiltonian is defined for the physical system plus thermostat variables 
which incorporates a coordinate-dependent time scaling of Poincar\'{e}-Sundman type \cite{Bond98,Benest02}.
Restricting the dynamics to a fixed value (zero) of the extended Hamiltonian results in the system variables
sampling their phase space according to, for example, the canonical density \cite{Bond99} (subject to the 
assumption of ergodicity.)  The Hamiltonian version of the isokinetic
thermostat is described in Section \ref{sec:setting} below.

A fundamental question concerning deterministic thermostats has to do with the 
effective ergodicity of the dynamics on the timescale of the simulation.  
If the dynamics is not effectively ergodic, then 
trajectory simulations will not generate the correct invariant measure \cite{Cornfeld82,Sturman06}.
It has long been recognized, for example,  that the dynamical system 
consisting of a single harmonic oscillator degree of freedom coupled to the NH thermostat
variable is not ergodic \cite{Legoll07}.
A large amount of effort has  been expended in attempts 
to design thermostats exhibiting dynamics more ergodic than the basic NH system  
\cite{Martyna92,Leimkuhler04a,Hunenberger05,Leimkuhler05a}.

The question of ergodicity in thermostats is conceptually closely related  to the problem of
statistical versus nonstatistical behavior in the (classical) theory of unimolecular reaction
rates \cite{Robinson72,Gilbert90,Baer96}.  
Broadly speaking, in this case one would like to know whether a molecule will
behave according to a statistical model such as RRKM theory, or whether it will exhibit significant deviations
from such a theory, ascribable to nonstatistical dynamics \cite{deleon81,Carpenter}.  
Such `nonstatisticality', which
can arise from a number of dynamical effects, 
is analogous to the failure of ergodicity in deterministic thermostats.

In recent years there have been a number of 
theoretical and computational advances in the application
of dynamical systems theory  \cite{Wiggins90,Wiggins92,Wiggins94} 
to study reaction dynamics and phase space structure in multimode models of 
molecular systems, and to probe the
dynamical origins of nonstatistical behavior \cite{wwju,WaalkensSchubertWiggins08}.  
The fundamental chemical concept 
of the \emph{transition state}, defined  as a surface 
of no return in phase space, has been successfully and rigorously generalized from the well-established 
2 degree of freedom case  \cite{Pechukas81}
to systems with $N \geq 3$ degrees of freedom \cite{WaalkensSchubertWiggins08}.
Moreover, dynamical indicators exist (determination of reactive phase space volume, 
behavior of the reactive flux) to diagnose nonstatistical behavior.

Despite their obvious potential relevance for the questions at issue, there has been relatively little work
applying the powerful techniques from modern dynamical systems theory,
in particular the theory of multidimensional Hamiltonian systems \cite{Wiggins92,Wiggins94}, 
to study the phase space structure of deterministic thermostats
\cite{Posch86,Posch97,Hoover98b,Hoover01a,Dettmann96,Morriss98,Leimkuhler05a,Legoll07}.
There appears to be considerable scope for application of these and other approaches 
\cite{Martens87,Laskar93,VelaArevalo01}
to the dynamics of deterministic thermostats.  

In the present paper we begin the development of a novel theoretical framework for 
the study of thermostat dynamics.  
Specifically, we describe how 
recently developed methods for the analysis of multimode Hamiltonian systems can be applied
to investigate the phase space structure of the isokinetic thermostat \cite{Morriss98}.

Although not as widely used as the Nos\'{e}-Hoover thermostat and its many variants, 
the non-Hamiltonian version of the isokinetic thermostat has been developed and applied to
several problems of chemical interest by Minary et al.\ \cite{Minary03a,Minary03b}.
In this thermostat, the particle momenta are subject to a nonholonomic constraint
that keeps the kinetic energy, hence temperature, constant.  The resulting dynamics
generates a canonical distribution in configuration space \cite{Morriss98}.
A Hamiltonian version of the isokinetic thermostat was given by Dettmann \cite{Dettmann96,Morriss98}, 
and this Hamiltonian formulation (see also \cite{Litniewski93,Morishita03}) 
is the point of departure for our investigation.

The Hamiltonian formulation of the isokinetic thermostat is presented in Section \ref{sec:setting}.  
The non-Hamiltonian 
equations of motion for a Hamiltonian system subject to the isokinetic constraint 
are shown to correspond to Hamiltonian dynamics at zero energy under an extended Hamiltonian whose
potential is obtained from the physical potential by exponentiation.
The extended Hamiltonian dynamics are therefore nonseparable
and potentially chaotic (ergodic), even though the physical Hamiltonian 
might be separable.
For the Hamiltonians  we consider the  physical potential exhibits a saddle of index one, 
as for the case of a bistable reaction profile coupled to one or more
transverse confining modes.  The bistable mode can play two distinct roles in the theory:
it can either be interpreted as a reaction coordinate of physical interest, or as a 
thermalizing thermostat mode \cite{Minary03a}.

Essential concepts concerning the phase space structure of multimode Hamiltonian systems,
especially the significance of Normally Hyperbolic Invariant Manifolds (NHIMs) and their role in
the phase space structure and reaction dynamics of multimode molecular systems 
with index one saddle, are briefly reviewed in Section \ref{sec:fix_energy}. 

In Section \ref{sec:hamiltonian_nonhamiltonian}  we show
that the extended Hamiltonian dynamical system satisfies the 
same conditions satisfied by the physical Hamiltonian that give rise to the 
phase space structures discussed in Section \ref{sec:fix_energy}.  
We then show that these phase space structures exist for the non-Hamiltonian isokinetic 
equations of motion for the thermostatted physical Hamiltonian by an explicit mapping.
Section \ref{sec:summary} concludes.

\section{The physical Hamiltonian, the extended Hamiltonian, 
and non-Hamiltonian isokinetic thermostat}
\label{sec:setting}

We begin with a  physical Hamiltonian of the standard form 
\begin{equation}
\label{ham_1}
H(q, p) = \frac{1}{2} p^2 + \Phi(q),
\end{equation}
with Hamilton's equations given by:
\begin{subequations}
\label{hameq_1}
\begin{align}
\dot{q} & = \frac{\partial H}{\partial p} = p. \\
\dot{p} & =  -\frac{\partial H}{\partial q} = -\Phi_q (q),
\end{align}
\end{subequations}
\noindent
where $(q, p) \in \mathbb{R}^n \times \mathbb{R}^n$  
are the physical coordinates and  $\Phi(q)$ is the potential energy. 
Following Dettmann and Morriss \cite{Dettmann96,Morriss98},
we construct a Hamiltonian system having the property 
that trajectories on a fixed energy surface of the new Hamiltonian 
correspond to trajectories of the physical Hamiltonian \eqref{ham_1} 
which satisfy an isokinetic constraint in the physical coordinates. 
An extended Hamiltonian $\calK$ is defined as follows:
\begin{equation}
\label{sp_3}
\calK(q, \pi) = e^{-\fb \Phi} \, \calH_{\fb} 
\end{equation}
\noindent
where $\calH_{\fb}$ is 
\begin{equation}
\label{H_beta}
\calH_{\fb} = \frac{1}{2} e^{(\fb + 1)\Phi} \pi^2 - \frac{1}{2} e^{(\fb - 1)\Phi},
\end{equation}
\noindent
$\fb$ is an arbitrary parameter, and the relation between momentum variables $p$ and $\pi$ is specified below.
The value chosen for the parameter $\fb$  defines a particular time scaling via factorization of $\calK$; 
setting $\fb=-1$, for example,  ensures that $\calH_{\fb}$ has $q$-independent kinetic energy.
(For simplicity we measure energies in units of $k_{\text{B}} T$, thus keeping the value of $T$ implicit.)

The Hamiltonian \eqref{sp_3} includes a time scaling factor $e^{-\fb\Phi}$, and
Hamilton's equations of motion for $\calK$ in {physical} time $t$ are:
\begin{subequations}
\label{sp_4}
\begin{align}
\dot{q} & = +\pd{\calK}{\pi} = e^{\Phi}\, \pi\\
\label{sp_4b}
\dot{\pi} & = -\pd{\calK}{q} = -\Phi_q  
\left[ \frac{1}{2} e^{\Phi} \pi^2 
+ \frac{1}{2} e^{-\Phi} \right]
\end{align}
\end{subequations}
\noindent
and are manifestly $\fb$-independent. 

To show that trajectories of the 
Hamiltonian system \eqref{sp_4} with $\calK = 0$ correspond to trajectories of 
the physical system \eqref{ham_1} satisfying the isokinetic constraint,  
first note that 
the time derivative of $\calH_{\fb}$  {\em along trajectories of \eqref{sp_4}} is given by
\begin{subequations}
\begin{align}
\dot{\calH}_{\fb} & = \dot{q} \, \pd{\calH_{\fb}}{q} + \dot{\pi} \, \pd{\calH_{\fb}}{\pi}\\
&= \fb \Phi_q e^{\Phi} \pi \calH_{\fb}.
\end{align}
\end{subequations}
\noindent
This implies that trajectories of \eqref{sp_4} satisfying $\calH_{\fb} = 0$ at $t = 0$,  
satisfy $\calH_{\fb}=0$  for all $t$
(for arbitrary  $\fb$).  Using \eqref{sp_3}, this implies that these 
trajectories are also confined to the surface
$\calK = 0$ for all time.

The relationship between the Hamiltonian dynamics of \eqref{sp_4} on $\calK = 0$  
and isokinetic trajectories of \eqref{ham_1} is made apparent by making the
\emph{noncanonical} transformation of variables
\begin{subequations}
\label{coord_trans}
\begin{align}
q & \mapsto q, \\
\pi & \mapsto   e^{-\Phi(q)} p. 
\end{align}
\end{subequations}
This coordinate transformation is clearly invertible and is, in
fact, a diffeomorphism (as differentiable as $\Phi$). Applying \eqref{coord_trans} to 
$\calH_{\fb}$ gives:
\begin{equation}
\calH_{\fb} =  \frac{1}{2}e^{(\fb - 1)\Phi}(p^2 -1 )
\end{equation}
\noindent
from which we can immediately conclude that  trajectories of the Hamiltonian 
system \eqref{sp_4} on  $\calK = \calH_{\fb} =0$ automatically satisfy the isokinetic condition
\begin{equation}
p^2 = 1,
\end{equation}
\noindent
in the physical coordinates $(q, p)$. 
Substituting relation \eqref{coord_trans} into \eqref{sp_4}
we obtain equations of motion for $(q, p)$:
\begin{subequations}
\label{isokinetic_1}
\begin{align}
\dot{q} & = p  \label{isokinetic_1a}\\
\dot{p} & =  - \Phi_q \, \frac{1}{2}(p^2 +1) + p (\Phi_q \cdot \dot{q}) = -\Phi_q - \alpha p  \label{isokinetic_1c}
\end{align}
\end{subequations}
\noindent
where $\alpha \equiv -\Phi_q \cdot p$ and we have used the constraint $p^2 = 1$.
Equations \eqref{isokinetic_1} are the isokinetic
equations of motion for the 
thermostatted  physical Hamiltonian \eqref{ham_1}
in physical time $t$, obtained via Gauss' principle of least constraint \cite{Morriss98}.

By design, the isokinetic dynamics \eqref{isokinetic_1} 
generates a canonical distribution in the coordinates
$q$ \cite{Dettmann96,Morriss98}.
Minary et al.\ \cite{Minary03a} have shown that the addition of
thermalizing degrees of freedom to the physical Hamiltonian \eqref{ham_1} can facilitate the
attainment of the correct canonical distribution in $q$-space.
If $H$ describes a collection of uncoupled oscillators, 
addition of a bistable thermalizing degree of freedom renders the
Hamiltonian dynamics under $\calK$ isomorphic to that of a reactive
degree of freedom coupled to several bath modes, so that
we can obtain useful insights into the thermostat dynamics using
methods recently developed for multidimensional Hamiltonian 
systems. Alternatively, if $H$ describes a reactive mode coupled to 
bath modes, then the $\calK$ dynamics is already in an appropriate form
for the phase space analysis described  in the next Section.

\section{Phase Space Structures on a Fixed Energy Surface}
\label{sec:fix_energy}

Our analysis of thermostat dynamics  will be carried out in phase space,
using the tools and framework for reaction type dynamics of Hamiltonian systems
developed in \cite{wwju,ujpyw,WaalkensBurbanksWiggins04,WaalkensWiggins04,
WaalkensBurbanksWigginsb04,WaalkensBurbanksWiggins05,WaalkensBurbanksWiggins05c,
SchubertWaalkensWiggins06,WaalkensSchubertWiggins08}. We will show in Section \ref{sec:hamiltonian_nonhamiltonian} that these results apply both to the physical Hamiltonian system \eqref{hameq_1} and the extended Hamiltonian system \eqref{sp_4}. Here we give a brief summary of
the setting and relevant results from these references.

The starting point for identifying a region of phase space relevant to reaction is to locate an
equilibrium point of Hamilton's equations, denoted $(q^{*}, p^{*})$, that is of
saddle-centre-$\ldots$-centre stability type. By this we mean that 
the matrix associated with the linearization of Hamilton's
equations about this equilibrium point has two real eigenvalues of
equal magnitude, with one positive and one negative, and $n-1$
purely imaginary complex conjugate pairs of eigenvalues. We will
assume that the purely imaginary eigenvalues satisfy a generic  nonresonance condition 
in the sense that they are independent over the rational numbers (this is discussed in more detail in Section \ref{sec:hamiltonian_nonhamiltonian}).

We will assume that such an equilibrium point is present in the physical system \eqref{hameq_1}
and we will show that the same type of equilibrium point exists for the extended Hamiltonian system \eqref{sp_4} in Section \ref{sec:hamiltonian_nonhamiltonian}. However, the discussion in this section applies to any type of Hamiltonian system near the same type of equilibrium point.  Without loss of generality we can assume that  $(q^{*}, p^{*})$ is  located at the origin, and we denote its energy by
$H(q^{*}, p^{*}) \equiv h^\ast$.

We will be concerned with geometrical structures in a neighborhood 
of phase space containing the saddle-centre-$\ldots$-centre type equilibrium point. 
We emphasize this fact by denoting the neighborhood by 
$\mathcal{L}$; this region is to be chosen so
that a new set of coordinates can be constructed (the normal form
coordinates) in which the Hamiltonian can be expressed (the normal
form Hamiltonian) such that it provides an integrable nonlinear
approximation to the dynamics which yields phase space structures
to within a given desired accuracy.

For $h-h^\ast$ sufficiently small and positive, locally the ($2n-1$ dimensional)
energy surface $\Sigma_h$ has the structure of
$S^{2n-2}\times\mathbb{R}$ in the $2n$-dimensional phase space.
The energy surface $\Sigma_h$ is split locally into two
components, ``reactants'' (R) and ``products'' (P), by a ($2n-2$
dimensional) ``dividing surface'' ($\text{DS}(h)$) that is
diffeomorphic to $S^{2n-2}$. The dividing surface that we
construct has the following properties:

\begin{itemize}

\item The only way that trajectories can evolve from reactants (R)
to products (P) (and vice-versa), without leaving the local region
$\mathcal{L}$, is through $\text{DS}(h)$. In other words, initial
conditions  on  this dividing surface  specify all reacting
trajectories.

\item The dividing surface is free of local
re-crossings; any trajectory which crosses it must leave the
neighbourhood $\mathcal{L}$ before it might possibly cross again.

\item The dividing surface  minimizes the (directional) flux.

\end{itemize}

The fundamental phase space building block that allows the
construction of a dividing surface with these properties is a
particular \emph{Normally Hyperbolic Invariant Manifold} (NHIM)
which, for fixed energy $h > h^\ast$, will be denoted
$\text{NHIM}(h)$.  The $\text{NHIM}(h)$ is diffeomorphic to $S^{2n-3}$ and
forms the natural \emph{dynamical equator} of the dividing
surface: The dividing surface is split by this equator into
$(2n-2)$-dimensional hemispheres, each diffeomorphic to the open
$(2n-2)$-ball, $B^{2n-2}$.  We will denote these hemispheres by
$\text{DS}_\text{f}(h)$ and $\text{DS}_\text{b}(h)$ and call them the
``forward reactive'' and ``backward reactive'' hemispheres,
respectively.  $\text{DS}_\text{f}(h)$ is crossed by trajectories
representing ``forward'' reactions (from reactants to products),
while $\text{DS}_\text{b}(h)$ is crossed by trajectories representing
``backward'' reactions (from products to reactants).

The $(2n-3)$-dimensional $\text{NHIM}(h)$ is an (unstable) invariant subsystem
which, in chemistry terminology, corresponds to the energy surface
of the ``activated complex'' \cite{Pechukas81,Truhlar1}.

The $\text{NHIM}(h)$ is of saddle stability type, having $(2n-2)$-dimensional
stable and unstable manifolds $W^s(h)$ and $W^u(h)$ that are
diffeomorphic to $S^{2n-3}\times\mathbb{R}$.  Being of
co-dimension one 
\cite{footnote1}  
with respect to the energy surface, these
invariant manifolds act as separatrices, partitioning the energy
surface into ``reacting'' and ``non-reacting'' parts.

These phase space structures can be computed via an algorithmic
procedure based on Poincar{\'e}-Birkhoff normalization \cite{wwju,ujpyw,WaalkensSchubertWiggins08}. 
This involves developing a new set of coordinates, the {\em normal form
coordinates}, $(\qb, \pb)$, which are realized through a  symplectic coordinate
transformation from the original, physical 
coordinates,
\begin{equation}
T(q, p) = (\qb, \pb),
\end{equation}
which, in a local neighbourhood $\mathcal{L}$ of the equilibrium point,
``unfolds'' the dynamics into a ``reaction coordinate'' and ``bath
modes''. Expressing $H$ in the new coordinates, $(\qb, \pb)$, via
\begin{equation}
H_\text{NF}(\qb, \pb) = H(T^{-1} (q, p)),
\end{equation}
\noindent gives $H_\text{NF}$ in a simplified form. 
The normalization procedure can also be adapted to yield explicit
expressions for the coordinate transformations, $T(q, p) = (\qb, \pb)$
and $T^{-1}(\qb, \pb) = (q, p)$, between the normal form (NF)
coordinates and the original coordinates 
\cite{footnote2}. 
These coordinate transformations are essential for physical
interpretation of the phase space structures that we construct in
normal form coordinates since they allow us to transform these
structures back into the original ``physical'' coordinates.

The nonresonance condition implies that the normal form procedure yields an explicit expression for the
normalised Hamiltonian $H_\text{NF}$ as a function of $n$ local
integrals of motion:
\begin{equation}
H_\text{NF} = H_\text{NF}(I_1, I_2, \ldots, I_n).
\end{equation}
\noindent The integral, $I_1$, corresponds to a ``reaction
coordinate'' (saddle-type DoF):
\begin{equation}
I_1 = \qb_1 \pb_1,
\end{equation}
\noindent The integrals $I_k$, for $k=2,\ldots,n$, correspond to
``bath modes'' (centre-type DoFs):
\begin{equation}
I_k = \frac{1}{2}\left(\qb_k^2 + \pb_k^2\right).
\end{equation}

The integrals provide a natural definition of the term ``mode''
that is appropriate in the context of reaction, and their existence is a
consequence of the (local) integrability in a neighborhood of the
equilibrium point of saddle-centre-$\ldots$-centre stability type.
Moreover, the expression of the normal form Hamiltonian in terms
of the integrals provides us a way to partition the energy between
the different modes \cite{footnote3}.

The $n$ integrals, the normalized Hamiltonian expressed as a function of the integrals, 
and the transformation between the normal form coordinates and the physical coordinates 
are the key to practically realizing the phase space structures described at the  beginning of this section.  
The approximate integrability of Hamilton's equations in the reaction region allows a 
precise and quantitative understanding of all possible trajectories in this region. 
It also provides a natural construction of an energy dependent reaction coordinate 
whose properties are determined solely by the Hamiltonian dynamics, 
as opposed to the need for  a priori definitions of possible candidates 
for reaction coordinates \cite{Heidrich95}.

The $n$ integrals of the motion defined in the neighborhood of the reaction region give 
rise to further phase space structures, and therefore constraints on the motion, 
in addition to those described at the beginning of this section.  
The common level sets of all the integrals are examples of invariant 
{\em Lagrangian submanifolds} \cite{Arnold78,Maslov81,Littlejohn92}, which have the geometrical structure
of two disjoint $n$-dimensional toroidal cylinders, denoted 
$\mathbb{R} \times \mathbb{T}^{n-1}$, i.e. the cartesian product of a line with $n-1$ copies of the circle.

In the next section we show how all of this phase space structure exists for 
thermostatted dynamics of the physical Hamiltonian \eqref{ham_1} in physical time $t$.

\section{
Microcanonical phase space structure: Hamiltonian and corresponding non-Hamiltonian thermostatted systems}
\label{sec:hamiltonian_nonhamiltonian}

In this section we will show that if the phase space structure described in Section 
\ref{sec:fix_energy}  exists for the physical Hamiltonian system \eqref{hameq_1}, 
it  also exists in the phase space of the non-Hamiltonian isokinetic
equations of motion \eqref{isokinetic_1} corresponding to the 
thermostatted dynamics of the physical Hamiltonian \eqref{ham_1}
in physical time $t$.  This is accomplished in 3 steps by showing:

\begin{enumerate}

\item If the physical Hamiltonian system \eqref{hameq_1} has an 
equilibrium point at the origin  of saddle-centre-$\ldots$-centre stability type,  
then the Hamiltonian system defined by \eqref{sp_4} corresponding to the Hamiltonian isokinetic 
thermostat has an equilibrium point at the origin also of saddle-centre-$\ldots$-centre stability type. Morever (and significantly), we show that the equilibrium points in these two systems satisfy the same non-resonance condition.

\item The energy of the saddle-centre-$\ldots$-centre type equilibrium point of \eqref{sp_4} is negative, 
but it can be brought sufficiently close to zero so that the microcanonical geometrical structures 
described in section \ref{sec:fix_energy} exist on the zero energy surface of \eqref{sp_4}.

\item The geometrical structures on the zero energy surface of \eqref{sp_4}
map to geometrical structures in the phase space of the non-Hamiltonian thermostatted 
system corresponding to \eqref{isokinetic_1}.

\end{enumerate}

\noindent
We begin with step 1.  We assume that \eqref{hameq_1} has an equilibrium point
at $(q,p) = (q^\ast, p^\ast) = (0,0)$. From \eqref{ham_1}, the energy of this equilibrium point is
$H(0,0) = \Phi(0)$.

The stability of the equilibrium point is determined by the
eigenvalues of the derivative of the Hamiltonian vector field evaluated at the equilibrium
point. This is given by the $2n \times 2n$ matrix:
\begin{equation}
\mbox{M}_{\mbox{\tiny sys}} = \left(
\begin{array}{cc}
0_{n \times n} & \mbox{id}_{n \times n} \\
-\Phi_{qq}(0) & 0_{n \times n}
\end{array}
\right), \label{Hess_ham}
\end{equation}
\noindent where $0_{n \times n}$ denotes the $n \times n$ matrix
of zeros and  $\mbox{id}_{n \times n}$ denotes the $n \times n$
identity matrix. We require the equilibrium point to be of
saddle-centre-$\ldots$-centre stability type. This means that the
$2n \times 2n$ matrix $\mbox{M}_{\mbox{\tiny sys}}$ has
eigenvalues $\pm \lambda, \pm i \omega_i$, $i=2, \ldots, n$ where
$\lambda$ and $\omega_i$ are real.

Eigenvalues $\gamma$ of $\mbox{M}_{\mbox{\tiny sys}}$ are obtained by solving
the characteristic equation $\det(\mbox{M}_{\mbox{\tiny sys}}  - \gamma \mbox{id}_{2n \times 2n}) = 0$. 
From Theorem 3 of \cite{silvester}, the block structure of the 
$2n \times 2n$ matrix $\mbox{M}_{\mbox{\tiny sys}}$ implies that 
\begin{equation}
\det(\mbox{M}_{\mbox{\tiny sys}}  - \gamma \mbox{id}_{2n \times 2n})
= \det(\Phi_{qq}(0) + \gamma^2 \mbox{id}_{n \times n})=0
\label{char_sys}
\end{equation}
\noindent
so that the $2n$ eigenvalues $\gamma$ are given in terms of $\sigma$, the eigenvalues
of the $n \times n$ Hessian matrix $\Phi_{qq} (0)$ associated with the potential as follows:
\begin{equation}
\gamma_{k}, \gamma_{k+n} = \pm \sqrt{-\sigma_k}, \;\;  k=1,\ldots, n.
\label{eig_phys}
\end{equation}
\noindent
Therefore, if $\Phi(q)$ has a rank-one saddle at $q=0$, so that one eigenvalue is strictly negative and
the rest are strictly positive, then $(q,p) = (0,0)$ is a
saddle-centre-$\ldots$-centre  type equilibrium point for \eqref{hameq_1} as described above.

We discuss the non-resonance condition in more detail. 
Suppose $\sigma_1<0$ and $\sigma_i >0$. $i=2, \ldots , n$.  T
hen the non-resonance condition satisfied by the purely imaginary eigenvalues 
is given by $(m_2, \ldots , m_n) \cdot (\gamma_2, \ldots , \gamma_n) \neq 0$  {\em for all} 
integer vectors $(m_2, \ldots , m_n)$ whose entries are not {\em all} zero (where ``$\cdot$'' 
denotes the scalar product). 
The non-resonance condition is responsible for the existence of the $n-1$ (local) integrals 
of motion $I_2, \ldots , I_n$.

Next, we consider the Hamiltonian system \eqref{sp_4} corresponding to the Hamiltonian 
isokinetic thermostat. It is easy to verify that $(q, \pi) = (0, 0)$ is an
equilibrium point for \eqref{sp_4} with energy ${\calK}
(0,0) = -\frac{1}{2} e^{-\Phi(0)}$.

Proceeding as above, we determine stability of this equilibrium point by computing the
matrix associated with the linearization of \eqref{sp_4} at the equilibrium point:
\begin{equation}
\mbox{M}_{\mbox{\tiny therm}} = \left(
\begin{array}{cc}
0_{n \times n} & \mbox{id}_{n \times n} e^{+\Phi(0)} \\
-\Phi_{qq}(0) \left(\frac{1}{2} e^{-\Phi(0)} \right) & 0_{n
\times n}
\end{array}
\right) \label{Hess_therm}
\end{equation}
The $2n$ eigenvalues of $\mbox{M}_{\mbox{\tiny therm}}$, which we  denote as
$\bar{\gamma}$, can be computed by exactly the same type of calculations as above. 
The resulting eigenvalues are given in terms of the eigenvalues of the potential Hessian as follows:
\begin{equation}
\bar{\gamma}_{k}, \bar{\gamma}_{k+n} = \pm \sqrt{-\frac{\sigma_k}{2}}, \;\;  k=1,\ldots, n.
\label{eig_therm}
\end{equation}
\noindent
Therefore,  it is clear that if  the potential of the physical Hamiltonian, $\Phi(q)$, has a rank-one saddle at $q=0$, 
so that one eigenvalue is strictly negative and
the rest are strictly positive, then $(q, \pi) = (0,0)$ is a
saddle-centre-$\ldots$-centre  type equilibrium point for \eqref{sp_4}. 

Moreover, since $\bar{\gamma} = \frac{1}{\sqrt{2}} \gamma$ it follows by comparing \eqref{eig_phys} with \eqref{eig_therm} that if 
the imaginary parts of the eigenvalues associated with the saddle for the 
physical Hamiltonian satisfy a non-resonance condition, then they satisfy a 
non-resonance condition for the saddle associated with the Hamiltonian isokinetic thermostat, i.e 
$(m_2, \ldots , m_n) \cdot (\gamma_2, \ldots , \gamma_n) \neq 0$ implies that 
 $(m_2, \ldots , m_n) \cdot \frac{1}{\sqrt{2}}(\gamma_2, \ldots , \gamma_n) = 
 (m_2, \ldots , m_n) \cdot (\bar{\gamma}_2, \ldots , \bar{\gamma}_n)\neq 0$.

Now consider step 2.  As we showed above, the saddle-centre-$\ldots$-centre type equilibrium point 
$(q, \pi)=(0,0)$  of \eqref{sp_4} has energy ${\cal K} (0,0) = -\frac{1}{2}
e^{-\Phi(0)}<0$. However, we are only interested in the dynamics on
the ${\calK}=0$ energy surface. The point here is that all of the
phase space structures described in section \ref{sec:fix_energy} exist for energies 
``above and sufficiently close'' to the energy  of the saddle-centre-$\ldots$-centre 
type equilibrium point, and the question is whether or not
$-\frac{1}{2} e^{-\Phi(0)}<0$ is close enough to zero so that the
phase space structures described in Section \ref{sec:fix_energy} 
exist on the ${\calK}=0$ energy surface.
This can easily be arranged by making $\Phi(0)$ larger by adding an appropriate constant to $\Phi(q)$, 
or by changing the value of the temperature $T$.

The final step 3 is to show that the phase space structure of
\eqref{sp_4} on ${\calK}=0$ exists for the isokinetic
equations of motion \eqref{isokinetic_1} corresponding to the 
thermostatted dynamics of the physical Hamiltonian \eqref{ham_1}
in physical time $t$.

Step 3 follows from general results that show that invariant manifolds and their 
stability properties are preserved under differentiable, invertible (with differentiable inverse) 
coordinate transformations (i.e. they are preserved under differentiable conjugacies). We emphasize that these results are ''well-known'' 
and appear in a variety of places throughout the literature, e.g., \cite{Llave92,flm}.

These general results allow us to make the following conclusions:
\begin{itemize}

\item Under the map \eqref{coord_trans} the $2n-1$ dimensional invariant energy surface 
of the effective Hamiltonian system \eqref{sp_4} maps to a $2n-1$ 
dimensional invariant  manifold for the non-Hamiltonian isokinetic
equations of motion for the 
thermostatted  physical Hamiltonian \eqref{ham_1}
in physical time $t$ defined by \eqref{isokinetic_1}.

\item Under the map \eqref{coord_trans} the $2n-3$ dimensional NHIM, 
its $2n-2$ dimensional stable and unstable manifolds, the $n$-dimensional invariant Lagrangian submanifolds, 
and the $2n-2$ dimensional dividing surface map to a $2n-3$ dimensional NHIM, 
its $2n-2$ dimensional stable and unstable manifolds, $n$-dimensional invariant submanifolds, 
and a $2n-2$ dimensional dividing surface in the $2n-1$ dimensional invariant  manifold in the $2n$ dimensional phase space of the non-Hamiltonian isokinetic
equations of motion for the 
thermostatted  physical Hamiltonian \eqref{ham_1}
in physical time $t$  defined by \eqref{isokinetic_1}.

\end{itemize}

\section{Summary and Outlook}
\label{sec:summary}

In this paper we have examined the relation between phase space
structures in Hamiltonian and non-Hamiltonian thermostats.
In particular, we have established the existence of a mapping
between invariant phase space structures in the phase space of the extended
Hamiltonian for the isokinetic thermostat and corresponding 
structures in the phase space of the non-Hamiltonian 
Gaussian isokinetic thermostat.  

Our results establish a conceptual link between 
the question of thermostat ergodicity and the issue of statisticality
in unimolecular isomerization reactions.
The existence of normally 
hyperbolic invariant manifolds in both the physical and extended
Hamiltonian phase spaces means that recently developed methods 
for the analysis of isomerization dynamics can be applied to
the thermostat problem.  Numerical studies based on the ideas presented here
are currently in progress.

Finally, we note that the approach presented here should be applicable to
other Hamiltonian thermostats, such as the Nos\'{e}-Poincar\'{e} system \cite{Bond99}.


\begin{acknowledgments}
Some of the original motivation for this work  comes from stimulating discussions on phase space 
structure in thermostatted systems that arose at the workshop on 
''Metastability and Rare Events in Complex Systems'' held at the 
Erwin Schr\"odinger Institute, Vienna, 18-22 Feb 2008, and organized by
Peter Bolhuis, Christoph Dellago and Eric Vanden-Eijnden.
SW  acknowledges the support of the  Office of Naval Research Grant No.~N00014-01-1-0769 and a useful correspondence with Rafael de la Llave concerning the nature of the preservation of phase space structures under maps.
\end{acknowledgments}


\begin{thebibliography}{69}
\bibitem[{foo({\natexlab{a}})}]{footnote1}
\bibinfo{note}{{Briefly, the co-dimension of a submanifold is the dimension of
  the space in which the submanifold exists, minus the dimension of the
  submanifold. The significance of a submanifold being ``co-dimension one'' is
  that it is one less dimension than the space in which it exists. Therefore it
  can ``divide'' the space and act as a separatrix, or barrier, to transport.}}
\bibitem[{foo({\natexlab{b}})}]{footnote2}
\bibinfo{note}{{The original coordinates $(q, p)$ had an interpretation as
  configuration space coordinates and momentum coordinates. The normal form
  coordinates $(\qb, \pb)$, in general, do not have such a physical
  interpretation since both $\qb$ and $\pb$ are nonlinear functions of both $q$
  and $p$.}}
\end{thebibliography}

\def\cprime{$'$} \def\cprime{$'$}

\end{document}